\documentclass[twocolumn,english]{iopart}
\usepackage[T1]{fontenc}
\usepackage{color}
\usepackage{textcomp}
\usepackage{mathrsfs}
\usepackage{relsize}
\usepackage{graphicx}
\usepackage{amssymb}

\makeatletter

\DeclareRobustCommand{\lyxmathsym}[1]{\ifmmode\begingroup\def\b@ld{bold}
  \def\rmorbf##1{\ifx\math@version\b@ld\textbf{##1}\else\textrm{##1}\fi}
  \mathchoice{\hbox{\rmorbf{#1}}}{\hbox{\rmorbf{#1}}}
  {\hbox{\smaller[2]\rmorbf{#1}}}{\hbox{\smaller[3]\rmorbf{#1}}}
  \endgroup\else#1\fi}

\usepackage{iopams}
\usepackage{setstack}

\makeatother

\usepackage{babel}

\begin{document}

\title{Calibration of Photomultiplier Arrays }

\author{F.~Neves$^{4,1}$, V.~Chepel$^{4}$, D.~Yu.~Akimov$^{5}$, H.~M.~Araújo$^{1,2}$,
E.~J.~Barnes$^{3}$, V.~ A.~Belov$^{5}$, A.~A.~Burenkov$^{5}$,
B.~Currie$^{1}$, B.~Edwards$^{1,2}$, C.~Ghag$^{3}$, M.~Horn$^{1}$,
A.~J.~Hughes$^{2}$, G.~E.~Kalmus$^{2}$, A.~S.~Kobyakin$^{5}$,
A.~G.~Kovalenko$^{5}$, V.~N.~Lebedenko$^{5}$, A.~Lindote$^{4}$,
M.~I.~Lopes$^{4}$, R.~Lüscher$^{3}$, K.~Lyons$^{1}$, P.~Majewski$^{2}$,
A.~StJ.~Murphy$^{3}$, J.~Pinto da Cunha$^{4}$, R.~Preece$^{3}$,
J.~J.~Quenby$^{1}$, P.~R.~Scovell$^{3}$, C.~Silva$^{4}$, V.~N.~Solovov$^{4}$,
N.~J.~T.~Smith$^{2}$, P.~F.~Smith$^{2}$, V.~N.~Stekhanov$^{5}$,
T.~J.~Sumner$^{1}$, C.~Thorne$^{1}$, R.~J.~Walker$^{1}$}

\address{$^{1}$Blackett Laboratory, Imperial College London, UK}

\address{$^{2}$Particle Physics Department, Rutherford Appleton Laboratory,
Chilton, UK}

\address{$^{3}$School of Physics and Astronomy, University of Edinburgh,
UK}

\address{$^{4}$LIP\textendash{}Coimbra \& Department of Physics of the University
of Coimbra, Portugal}

\address{$^{5}$Institute for Theoretical and Experimental Physics, Moscow,
Russia}
\begin{abstract}
A method is described that allows calibration and assessment of the
linearity of response of an array of photomultiplier tubes. The method
does not require knowledge of the photomultiplier single photoelectron
response model and uses science data directly, thus eliminating the
need for dedicated data sets. In this manner all photomultiplier working
conditions (e.g. temperature, external fields, etc) are exactly matched
between calibration and science acquisitions. This is of particular
importance in low background experiments such as ZEPLIN-III, where
methods involving the use of external light sources for calibration
are severely constrained.
\end{abstract}


\maketitle

\section{Introduction\label{sec:Introduction}}

Traditional procedures to characterize the response of a photomultiplier
tube (PMT) rely, typically, on the use of calibration light sources
and dedicated trigger setups. However, in some experiments, such as
the ZEPLIN-III WIMP search \cite{araujo:2006aa,Akimov:2007aa,Lebedenko:2008aa},
the use and positioning of these light sources is severely constrained
both by the low radioactivity background requirement and by the use
of VUV-rated components. Also, it is known that the response of a
PMT depends on its working conditions, namely, external fields and
temperature. In its already long history there have been several attempts
to model the response function of a PMT ($\mathscr{R}$). Nevertheless,
a general solution which covers different working conditions and different
types of PMT is still missing.

Based on experimental data obtained using Ni and Be dynodes, Wright
stated that the number of secondary electrons ejected per primary
electron is described by a Poisson distribution ($P$) \cite{Wright:1954ab}.
The effect of non-uniform photocathode and dynode surfaces or inter-stage
collection efficiencies (focusing optics) is the variation of the
mean of the distribution ($\eta$) from one primary electron to another,
thus increasing the variance of the PMT response function ($\sigma_{\mathscr{R}}$).
The calculations presented did not allow to conclude on the shape
of this response function, but only to infer that, being the effect
of non-uniformities negligible, the dominant statistics would be\textit{
gaussian} for a sufficiently large number of photoelectrons. Nevertheless,
Breitenberger \cite{Breitenberger:1955aa} reported that the electron
multiplication variance measured using activated BaO~+~SrO dynodes
is in fact larger than calculated when assuming a \textit{poissonian}
secondary emission process $P\left(n,\eta\right)$. Based on the same
assumption, Lombard \textit{et al.} \cite{Lombard:1961aa} derived
the pulse height spectrum for cascades starting with single photoelectrons.
The authors remarked that their results were inconsistent with observed
data, thus rejecting the hypothesis of the Poisson distribution ($P$)
being a good descriptor for the PMT electron multiplication process.
In spite of this conclusion, other authors\textcolor{blue}{ }\cite{Delaney:1964aa,Iinuma:1962aa,Tusting:1962aa}
consistently reported measurements which did agree with the calculations
by Lombard \textit{et al.} \cite{Lombard:1961aa} and attribute the
discrepant results of other work to noise in their experimental setup
\cite{Tusting:1962aa}. Using an \textit{exponential} distribution
to describe the electron multiplication at the dynodes, Prescott \textit{et
al.} \cite{Prescott:1962aa} obtained good agreement between calculated
and measured spectra for some specific types of PMT.

Baldwin \textit{et al}. \cite{baldwin:1965aa} suggested that the
inconsistent results mentioned above could be explained in terms of
the microscopic characteristics of the dynodes used. In fact, the
random orientations of the polycrystals in the Ag~+~MgO dynodes
used by Lombard \textit{et al.} \cite{Lombard:1961aa} are consistent
with a variation of the mean number of secondary electrons ($\eta$)
produced by primaries hitting different regions \cite{Wright:1954ab}.
On the other hand, the Sb~+~CsO dynodes used by Tusting \textit{et
al.} \cite{Tusting:1962aa} consist of a more uniform thin layer of
adsorbed material, which may account for a more constant $\eta$ across
the surface \cite{baldwin:1965aa}. A possible conclusion from these
evidences \cite{Gale:1966aa} is that one can assume that at each
PMT stage the electron multiplication process follows indeed a Poisson
distribution ($P$) given dynodes with uniform emission properties
\cite{Wright:1954ab,Breitenberger:1955aa,Lombard:1961aa}. 

In order to generalize the description of the fluctuations in the
secondary electron emission process, Prescott \cite{Prescott:1966aa}
had proposed the use of the \textit{Polya} distribution which contains
the \textit{Furry} (\textit{exponential}) and Poisson distributions
as special cases. The \textit{Polya} distribution is also used in
the description of cosmic-ray fluctuations and of charge multiplication
in proportional counters \cite{Prescott:1966aa}. For a PMT, the \textit{Polya}
describes the multiplication process when the number of secondary
electrons follows a Poisson distribution ($P$) with $\eta$ varying
across the dynode surface in a manner described by the \textit{Laplace}
distribution \cite{Prescott:1966aa}. Once again it was verified experimentally
that the \textit{Polya} distribution can only model the response for
a limited number of PMTs \cite{Prescott:1966aa,Wright:1987aa}.

More recent work involved a Monte Carlo simulation of dynode statistics
to assess the overall PMT response resolution\textcolor{blue}{ }\cite{Wright:1987aa}.
In spite of predicting accurately the resolution for a range of PMTs
\cite{Wright:1987aa}, the method demands the single electron response
(SER) to be measured experimentally. The issue is again that measuring
the SER at the working conditions (e.g. temperature, external field)
of PMTs installed in some experimental setups may present an insurmountable
challenge.

In the present work we propose an application of an existing method
to calibrate a PMT which does not demand the knowledge of its SER
model. Instead, the method relies only on the statistical description
of the light pulses arriving at the PMT photocathode. This fact eliminates
the need for dedicated data sets acquired using calibration light
sources, but allows for the use of the light pulses produced in the
sensitive volume of a detector during its science exposure. In this
manner, all the working conditions of a PMT (external fields, temperature,
light intensity, trigger, signal amplification and conditioning) are
perfectly matched between the calibration procedure and the science
data. One more significant advantage is that it eliminates any difficulties
posed by setting up the calibration light(s) in the context of a particular
experiment. Finally, one must emphasize that the proposed method of
calibration is more suitable for a detector having an array of PMTs
instead of a single one. This is related to the fact that an array
permits the implementation of some sort of position reconstruction,
thus allowing the effect of differences in the light collection efficiency
across the active volume of the detector to be minimised.

\section{Setup and Data Processing\label{sec:Setup}}

ZEPLIN-III is a two-phase (liquid/gas) xenon time projection chamber
designed to search for dark matter WIMPs \cite{araujo:2006aa,Akimov:2007aa,Lebedenko:2008aa}.
The active volume contains $\approx12$~kg of liquid xenon above
a compact hexagonal array of $31$ $2$-inch PMTs (ETL D730/9829Q).
The PMTs are immersed directly in the liquid at a temperature of $\approx-105$~$^{o}$C
and record both the rapid scintillation signal (S1) and a delayed
second signal (S2) produced by proportional electroluminescence in
the gas phase from charge drifted out of the liquid \cite{araujo:2006aa}.
The electric field in the active xenon volume is defined by a cathode
wire grid $36$~mm below the liquid surface and an anode plate $4$~mm
above the surface in the gas phase. These two electrodes define a
drift field in the liquid of $3.9$~kV/cm and an electroluminescence
field in the gas of $7.8$~kV/cm. A second wire grid is located $5$~mm
below the cathode grid just above the PMT array. This grid defines
a reverse field region which suppresses the collection of ionization
charge for events just above the array and helps to isolate the PMTs
input optics from the external high electric field.

The PMT signals are digitized at $2$~ns sampling over a time segment
of $\pm18$~$\mu$s either side of the trigger point. Each PMT signal
is fed into two $8$-bit digitizers (ACQIRIS DC265) with a $\times10$
gain diff{}erence between them provided by fast amplifi{}ers (Phillips
Scientifi{}c 770), to obtain both high and low sensitivity readout
covering a wide dynamic range. The PMT array is operated from a common
HV supply with attenuators (Phillips Scientifi{}c 804) used to normalize
their individual gains. The trigger is created from the shaped sum
signal of all the PMTs.

The raw data are processed and reduced by a purpose developed software
tool (ZE3RA), which finds candidate pulses in individual waveforms
by searching for signal excursions over a defined threshold ($V_{thr}$)
\cite{Lebedenko:2008aa}. Subsequent waveform processing includes
resolving adjacent/overlapping pulses, grouping of statistically consistent
structures (e.g. scintillation tails) and coincidence analysis of
occurrences in different channels. By design, ZE3RA outputs only amplitude,
area and timing parameters and does not ascribe any physical meaning
to pulses. This task is left to an independent software tool which
processes the original parameters assigning a physical meaning to
the reduced data. This assignment is made according to a well defi{}ned
set of rules, e.g. primary scintillation signals (S1) are fast and
must precede wider electroluminescence signals (S2).

Using S2 pulses from a $^{57}$Co source located above the instrument,
an iterative procedure was used to normalize the measured response
from each PMT (i.e. \textquoteleft{}fl{}at-fi{}eld\textquoteright{}
the array) \cite{Solovov:2008aa}. The procedure is based on fitting
to each channel a common cylindrical response profile extending away
from the vertical PMT axis and does not depend on the characterization
of the individual PMT response. Position reconstruction in the horizontal
plane was then achieved by using the converged response profi{}les
in a simultaneous least-squares minimization to all channels \cite{Solovov:2008aa}.
The vertical position is obtained by measuring the time difference
between S1 and S2 signals corresponding to the electron drift time
in the liquid.

\section{Methodology\label{sec:Methodology}}

Arising from the fact that photons follow Bose-Einstein statistics,
the Poisson distribution is a good approximation to the number of
photons arriving at the photocathode within a defined time window
\cite{fried:1965aa,Morton:1968aa}\textcolor{blue}{.} As the photoemission
process follows the binomial distribution (with the quantum efficiency
$\epsilon$ quantifying the probability of one photon producing one
photoelectron), it has been shown that the number of photoelectrons
$n$ produced in the photocathode also follows a Poisson distribution
\cite{fried:1965aa}\begin{equation}
P\left(\mu,n\right)=\frac{\mu^{n}e^{-\mu}}{n!}\ ,\label{eq:Poisson}\end{equation}
where $\mu$ is the mean number of photoelectrons. The value of $\mu$
has a simple relation to the mean number of incident photons of $\mu/\epsilon$.
Reworking Eq.~\ref{eq:Poisson} one obtains \cite{Dossi:2000aa}

\begin{equation}
\frac{P\left(\mu,0\right)}{\sum_{k=0}^{+\infty}P\left(\mu,k\right)}=\frac{N_{0}}{N}\Longrightarrow\mu=-ln\left(N_{0}/N\right)\ ,\label{eq:mu}\end{equation}
where $N$ stands for the total number of opened time windows (incident
light pulses) and $N_{0}$ for the number of times there were \textit{no
}photoelectrons produced in the photocathode.

In a general setup, the signals from a PMT are fed into some sort
of acquisition system (DAQ) allowing ultimately measurement of the
number of electrons arriving at the anode. This implies that the assertion
of the \textit{null }photoelectron signal ($P(\mu,0)$) must be made
against a measure of the noise intrinsic to the DAQ system used. In
ZEPLIN-III the noise distribution was parametrized using the same
waveforms containing the actual PMT signals \cite{Lebedenko:2008aa}.
To avoid any bias due to the occurrence of a transient or small signal,
the parametrization method relies on a consistency check of the noise
distribution variance during a sufficiently large time window. For
that purpose, the DAQ \textit{pre}-trigger region is divided into
$i=1..M_{0}$ consecutive regions containing $m$ samples each. For
each of these regions, the variance $\left\{ \sigma_{i}^{2},\, i=1..M_{0}\right\} $
of the signal amplitude distribution is calculated and the F-distribution
probability function ($Q$) is used to check if they are statistically
consistent:\begin{equation}
Q=\frac{\Gamma(\frac{\nu_{a}}{2}+\frac{\nu_{b}}{2})}{\Gamma(\frac{\nu_{a}}{2})\Gamma(\frac{\nu_{b}}{2})}\int_{0}^{\frac{\nu_{b}/2}{\nu_{b}/2+F\nu_{a}/2}}t^{\frac{\nu_{a}}{2}-1}(1-t)^{\frac{\nu_{b}}{2}-1}dt\ ,\label{eq:Q}\end{equation}
where

\[
\left\{ \begin{array}{ll}
\nu_{a}=m_{i}-1,\;\nu_{b}=m_{i+1}-1, & \sigma_{i}>\sigma_{i+1}\\
\nu_{a}=m_{i+1}-1,\;\nu_{b}=m_{i}-1, & \sigma_{i}\leq\sigma_{i+1}\end{array}\right.\ ,\]
and

\[
\left\{ \begin{array}{ll}
F=\sigma_{i}^{2}/\sigma_{i+1}^{2}, & \sigma_{i}>\sigma_{i+1}\\
F=\sigma_{i+1}^{2}/\sigma_{i}^{2}, & \sigma_{i}\leq\sigma_{i+1}\end{array}\right.\ .\]
$Q$ is therefore the significance level at which that hypotheses
($\sigma_{i}^{2}\equiv\sigma_{i+1}^{2},\, i=1..M_{0}$) can be rejected
\cite{Recipes:2002aa}. In the present work the values of $m=25$
($50$~ns) and $Q=0.0001$ were used. The maximum length of the total
sampled waveform was $2$~$\mu$s ($M_{0}=40$). The noise characterizing
each waveform is then defined as \begin{equation}
\sigma=\left\langle \sigma_{i}\right\rangle ,\, i=1..M\ ,\label{eq:sigma}\end{equation}
for those $M$ regions satisfying Eq.~\ref{eq:Q}. Waveforms for
which $M<20$ ($1$~$\mu$s) were not considered for the analysis.
Fig.~\ref{fig:noise_PMT1} shows the distribution of $\sigma$ values
for the central PMT in the ZEPLIN-III array. It can be seen that there
are two peaks both having a \textit{gaussian-}like profile. The peak
corresponding to higher values of $\sigma$ is due to the occurrence
of an external frequency pickup which can be identified by simple
visual inspection of the waveforms. With the described analysis this
presents no problem as the noise is parametrized independently for
each of the individual waveforms. 

\begin{center}
\begin{figure}[p]
\begin{centering}
\textsf{\includegraphics[width=9.5cm]{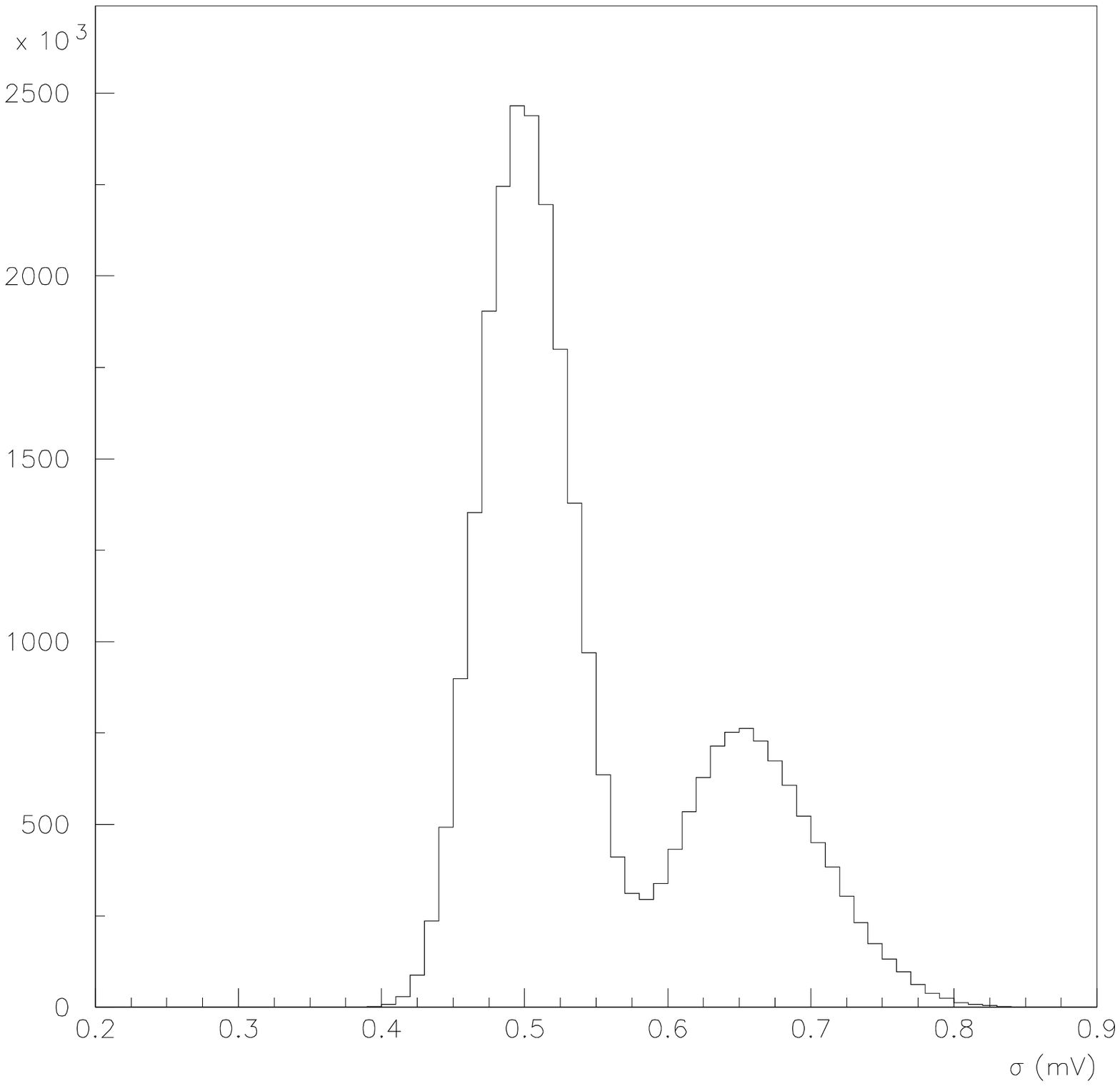}}
\par\end{centering}

\caption{\label{fig:noise_PMT1}Distribution of values of $\sigma$ (Eq.~\ref{eq:sigma})
for the central PMT in the ZEPLIN-III array.}

\end{figure}

\par\end{center}

Setting the software amplitude threshold ($V_{thr}$) to a certain
level ($k$) above the waveform noise ($\sigma$)

\begin{equation}
V_{thr}=k\sigma\ ,\label{eq:Threshold}\end{equation}
and selecting pulses which are predicted to have the same average
number of photons arriving at the photocathode of a particular PMT,
one can calculate $\mu$ (Eq.~\ref{eq:mu}) just by defining $N$
as the number of selected pulses and $N_{0}$ as the number of those
having an amplitude $V<V_{thr}$. This definition is the core of the
calibration method described here as it sets the conditions for observing
\textit{no} photoelectrons ($n=0$) or any number of photoelectrons
($n\geq1$) produced at the photocathode. Repeating the procedure
for all PMTs and a range of expected signal allows comparison of the
average PMT response in each iteration against the expected Poisson
mean ($\mu$). When selecting pulses, care must be taken to ensure
that $N-N_{o}\ggg N_{noise}$, where $N_{noise}$ is the expected
number of occurrences leaking from the noise distribution above $V_{thr}$.
For $k=3$ (Eq.~\ref{eq:Threshold}) and a normally distributed noise,
values of $\mu\gtrsim0.1$ should be used ($N\gtrsim1.13N_{0}$).
With this assumption the dominant error is the statistical uncertainty
associated with the Bernoulli trial of observing either $n=0$ or
$n\geq1$ photoelectrons from each incident light pulse. Defining
$B$ as the probability of $n=0$ at a given $V_{thr}$, the respective
variance from the Bernoulli distribution is expressed as $\sigma_{n=0}^{2}=B(1-B)$.
Applying the central limit theorem to a set of $N$ independent trials
(or incident light pulses), the variance of the random variable $N_{0}$
counting the number of $n=0$ occurrences can be written as \cite{Dossi:2000aa}

\begin{equation}
\sigma_{N_{0}}^{2}\cong N\sigma_{n=0}^{2}=NB(1-B)\ .\label{eq:sig_No}\end{equation}
Propagating this result into Eq.~\ref{eq:mu} we obtain \begin{equation}
\sigma_{\mu}^{2}\cong\frac{1-B}{NB}\ .\label{eq:sig_mu_B}\end{equation}
Considering that $N_{0}$ is drawn from a binomial distribution with
mean $NB$ then, taking the same validity constrains as for Eq.~\ref{eq:sig_No},
$N_{0}\cong NB$; feeding this into Eq.~\ref{eq:sig_mu_B} results
in

\begin{equation}
\sigma_{\mu}^{2}\cong\frac{1}{N_{0}}-\frac{1}{N}\ .\label{eq:sig_mu}\end{equation}
Combining Eq.~\ref{eq:mu} and Eq.~\ref{eq:sig_mu} one can derive
the number of incident light pulses ($N$) needed to keep the relative
error ($\delta$) below a required value\begin{equation}
\frac{\sigma_{\mu}^{2}}{\mu^{2}}<\delta^{2}\Longrightarrow N>\frac{1}{\mu^{2}\delta^{2}}\left(e^{\mu}-1\right)\ .\label{eq:sig_relative}\end{equation}
Whenever a limited statistics ($N$) is available, Eq.~\ref{eq:sig_relative}
can also be used to determine the interval for which $\mu$ can be
obtained within a certain accuracy $(\delta)$. 

Assuming that the photoelectron emission at the photocathode and the
secondary electron multiplication at the dynodes are independent,
the relative variance of the PMT response function ($\mathscr{R}$)
for $V>V_{thr}$ ($n>0$) can be obtained by adding the relative variances
from the distributions describing both processes

\begin{equation}
\left(\frac{\sigma_{\mathscr{R}}}{\left\langle \mathscr{R}\right\rangle }\right)_{V>V_{thr}}^{2}=\left(\frac{\sigma_{n>0}}{\mu_{n>0}}\right)^{2}+\frac{1}{\mu_{n>0}}\left(\frac{\sigma_{\mathscr{R}}}{\left\langle \mathscr{R}\right\rangle }\right)_{SER}^{2}\,,\label{eq:Rel_Var}\end{equation}
where \begin{equation}
\mu_{n>0}=\frac{\sum_{n=1}^{\infty}P(\mu,n)n}{\sum_{n=1}^{\infty}P(\mu,n)}=\frac{\mu}{1-e^{-\mu}}\,,\label{eq:mu_1}\end{equation}
and

\begin{equation}
\sigma_{n>0}^{2}=\frac{\sum_{n=1}^{\infty}P(\mu,n)(n-\mu_{n>0})^{2}}{\sum_{n=1}^{\infty}P(\mu,n)}\,,\label{eq:sigma_1}\end{equation}
represent, respectively, the mean ($\mu_{n>0}$) and the variance
($\sigma_{n>0}^{2}$) of the photoelectron distribution (Eq.~\ref{eq:Poisson})
for $n>0$. Using Eq.~\ref{eq:mu_1} and Eq.~\ref{eq:sigma_1},
the relative variance contribution from the photoelectron emission
process can be written as

\begin{equation}
\left(\frac{\sigma_{n>0}}{\mu_{n>0}}\right)^{2}=\frac{1-e^{-\mu}-\mu e^{-\mu}}{\mu}\,.\label{eq:Rel_Var_Poisson}\end{equation}
The contribution from the electron multiplication process in Eq.~\ref{eq:Rel_Var}
is derived simply by applying the central limit theorem to the PMT
SER relative variance ($(\sigma_{\mathscr{R}}/\left\langle \mathscr{R}\right\rangle )\lyxmathsym{\texttwosuperior}$)
when a set of $\mu_{n>0}$ photoelectrons are produced at the photocathode.

\section{Results\label{sec:Results}}

The following results were obtained using three different data sets,
which are described in detail in Ref.~\cite{Lebedenko:2008aa}:
\begin{enumerate}
\item low-energy Compton-scattered $\gamma$ events from a $^{137}$Cs calibration
source positioned above the detector;
\item low-energy events from a Am-Be neutron source positioned off-center,
above the detector;
\item $847$~kg.days of WIMP-search data acquired over $83$ days of continuous
stable operation. 
\end{enumerate}
The raw data were processed using a software threshold of $V_{thr}=3\sigma$
(Eq.~\ref{eq:Threshold}). The PMT calibration was performed using
the fast S1 (primary scintillation) signals. The expected number of
S1 photons arriving at each PMT photocathode for individual events
is derived from the $3D$ position reconstruction algorithm used.
The $\mu$ value is calculated for each PMT channel and for each range
of number of photons by applying the method described in Sec.~\ref{sec:Methodology}.
For each range the corresponding PMT response was calculated averaging
the area ($A$) of the selected pulses. The resulting $A(\mu)$ distributions
for the different PMTs and data sets were then fitted using a linear
function. The errors associated with the calculation of $A$ are insignificant
and therefore were not considered in the fitting procedure. The results
obtained show that there is a good linearity of response for all the
PMTs in the $0.2\lesssim\mu\lesssim4$ interval ($\delta\lesssim5\%$,
Eq.~\ref{eq:sig_relative}). The slope on the fitted lines is an
estimator for the mean response to the PMTs for single photoelectron
signals ($\left\langle \mathscr{R}\right\rangle _{SER}$). Both the
distributions of $\mu$ as a function of $A$ and the corresponding
fits for each of the data sets are shown in Figs.~\ref{fig:PMT_24}-\ref{fig:PMT_13}
for three different PMTs. These are representative of the results
found for the whole set of $31$ PMTs.

\begin{center}
\begin{figure}[p]
\begin{centering}
\includegraphics[width=9.5cm]{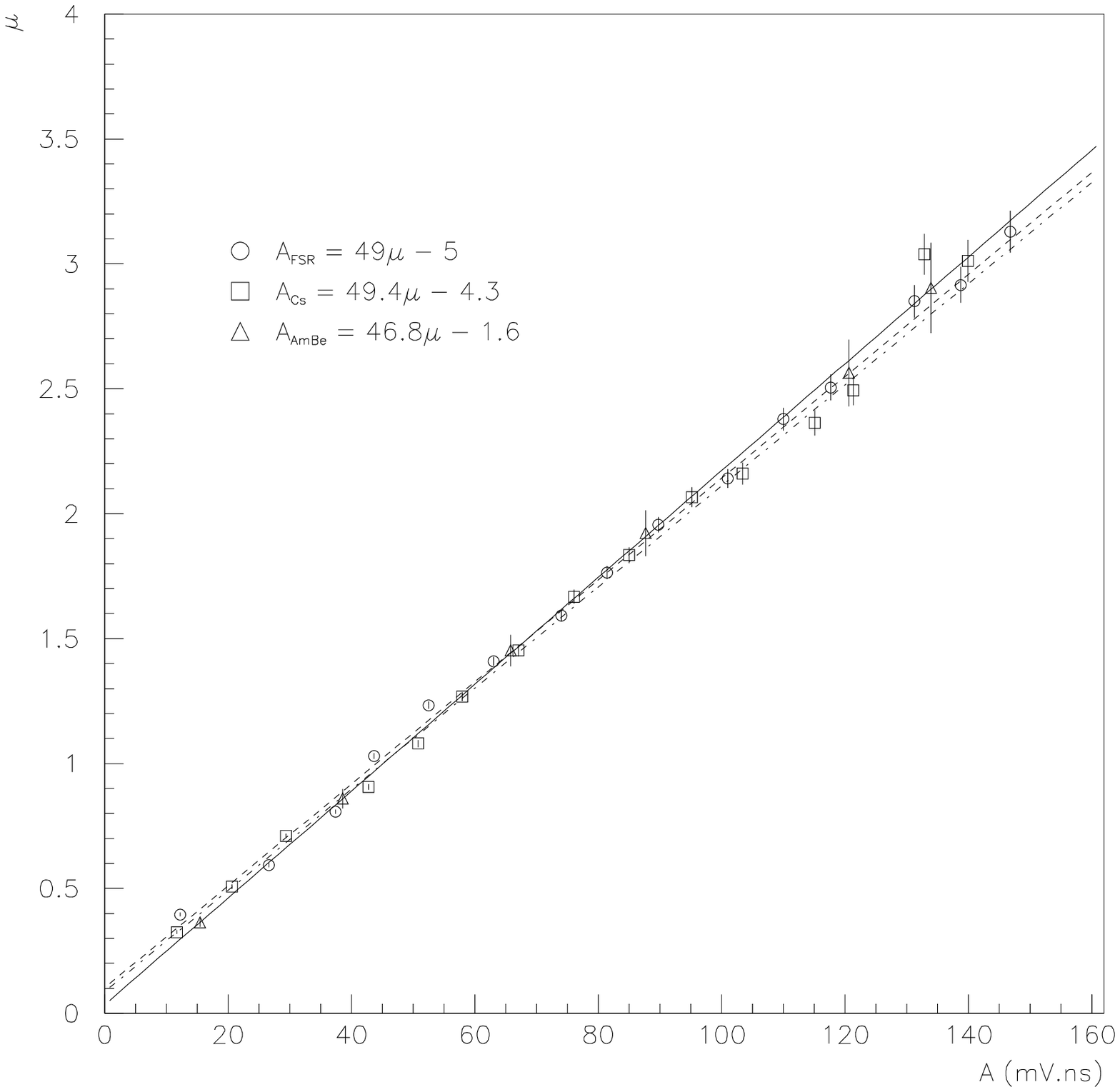}
\par\end{centering}

\caption{\label{fig:PMT_24}Calibration results for PMT $24$ in the ZEPLIN-III
array. The values of the $\mu(A)$ distributions and the corresponding
linear fits are shown for: (squares) $^{137}$Cs data set, (triangles)
Am-Be data set and (circles) WIMP-search data set.}

\end{figure}

\par\end{center}

\begin{center}
\begin{figure}[p]
\begin{centering}
\includegraphics[width=9.5cm]{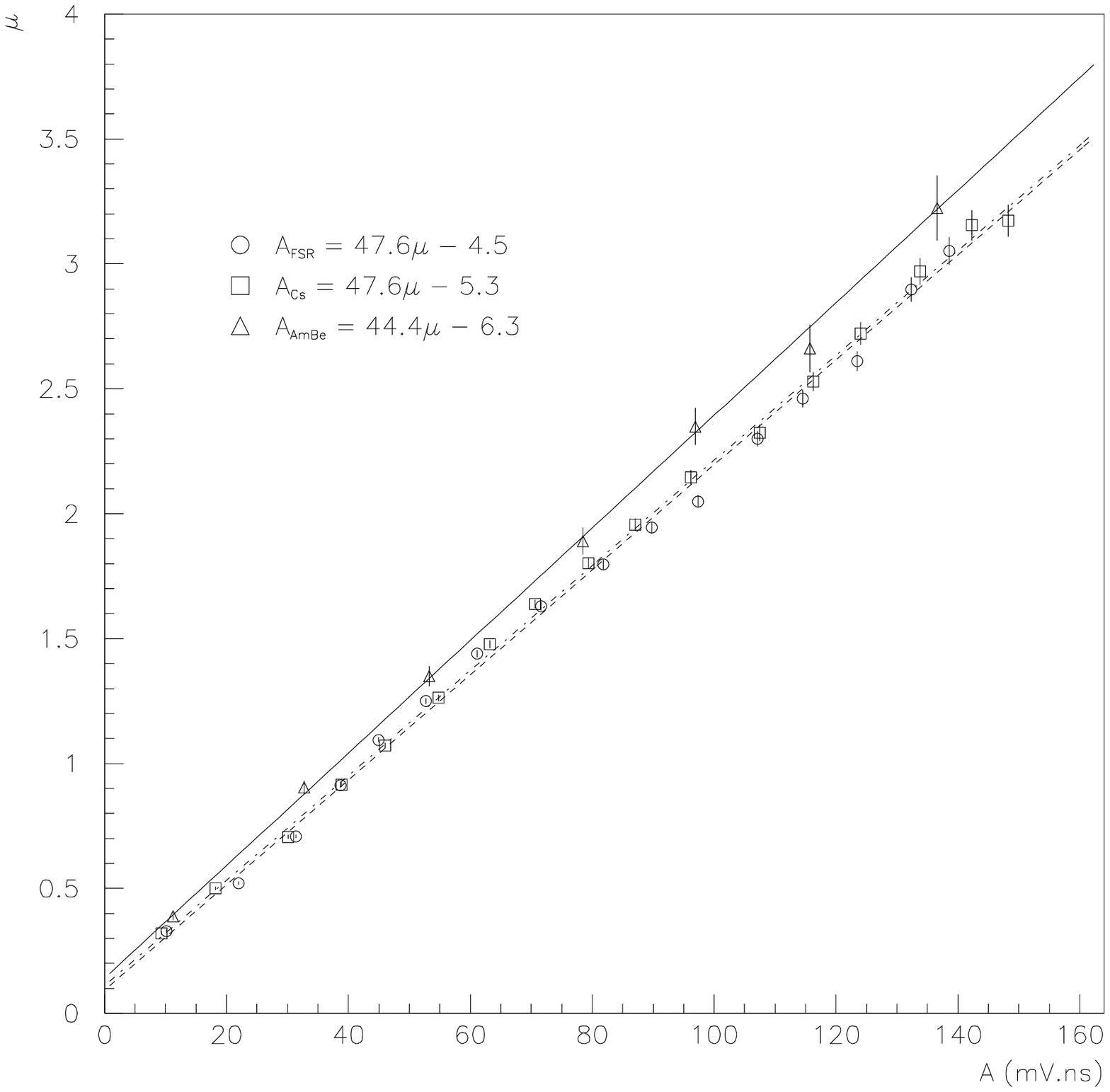}
\par\end{centering}

\caption{\label{fig:PMT_7}Calibration results for PMT $7$ in the ZEPLIN-III
array. The values of the $\mu(A)$ distributions and the corresponding
linear fits are shown for: (squares) $^{137}$Cs data set, (triangles)
Am-Be data set and (circles) WIMP-search data set.}

\end{figure}

\par\end{center}

\begin{center}
\begin{figure}[p]
\begin{centering}
\includegraphics[width=9.5cm]{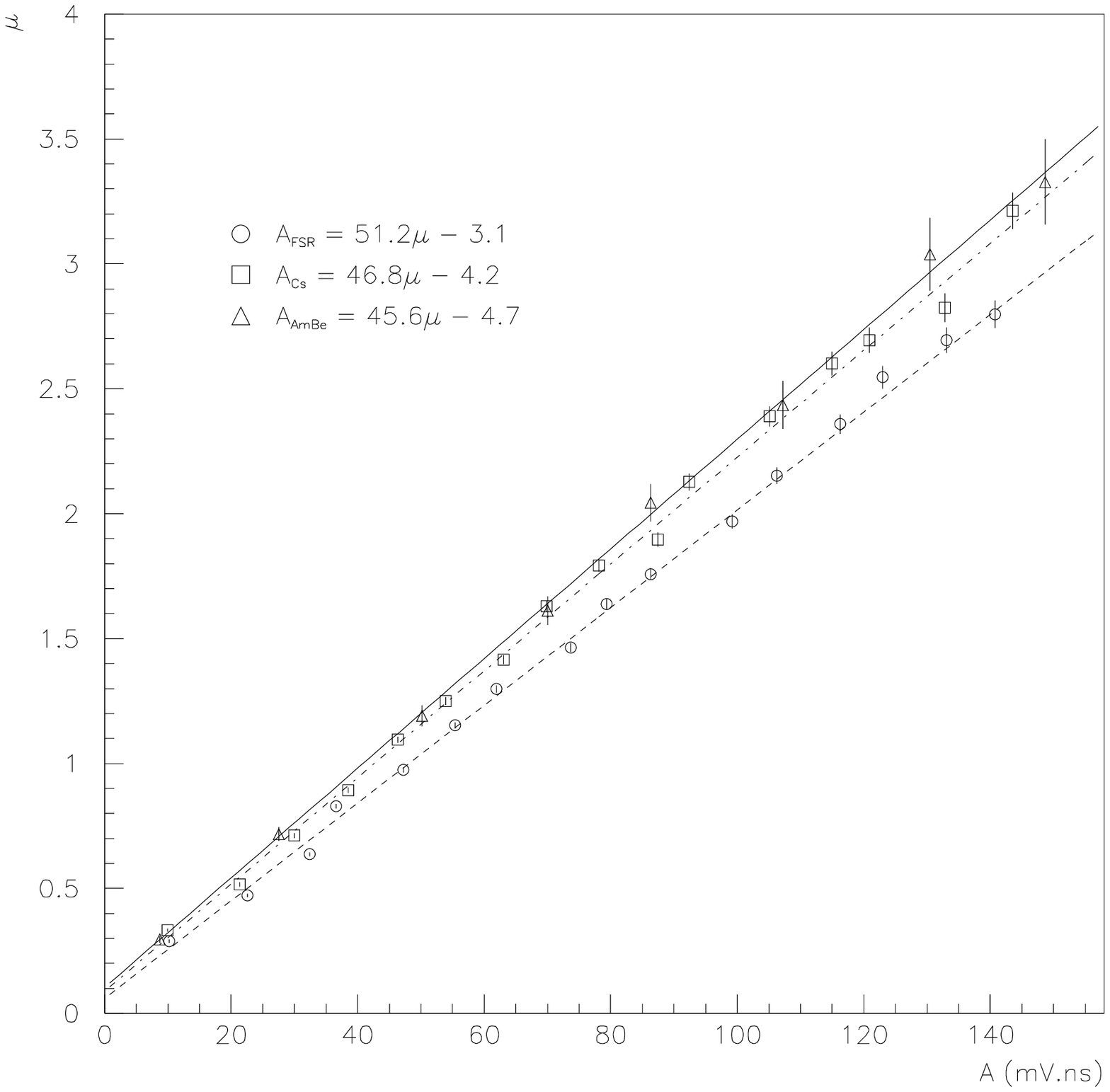}
\par\end{centering}

\caption{\label{fig:PMT_13}Calibration results for PMT $13$ in the ZEPLIN-III
array. The values of the $\mu(A)$ distributions and the corresponding
linear fits are shown for: (squares) $^{137}$Cs data set, (triangles)
Am-Be data set and (circles) WIMP-search data set.}

\end{figure}

\par\end{center}

The widths of the single electron response ($\left(\sigma_{\mathscr{R}}\right)_{SER}$)
for all the PMTs were determined feeding the estimated values of $\left\langle \mathscr{R}\right\rangle _{SER}$
into Eq.~\ref{eq:Rel_Var}. For each value of $\mu$, the relative
variance of the PMT response was calculated using the mean ($\left(\left\langle \mathscr{R}\right\rangle \equiv A\right)_{V>V_{thr}}$)
and \textit{root mean square} ($\left(\sigma_{\mathscr{R}}\equiv rms\right)_{V>V_{thr}}$)
from the corresponding area distribution of pulses above the threshold
($V>V_{thr}$, Eq.~\ref{eq:Threshold}). The errors concerning the
calculation of $A_{V>V_{thr}}$ and $rms_{V>V_{thr}}$ are insignificant
and were not considered. The obtained values of $\left(\sigma_{\mathscr{R}}\right)_{SER}$
for the $^{137}$Cs, Am-Be and WIMP-search data sets are shown in
Fig.~\ref{fig:PMT_7_RMS} for the same PMT also represented in Fig.~\ref{fig:PMT_7}. 

\begin{center}
\begin{figure}[p]
\begin{centering}
\includegraphics[width=9.5cm]{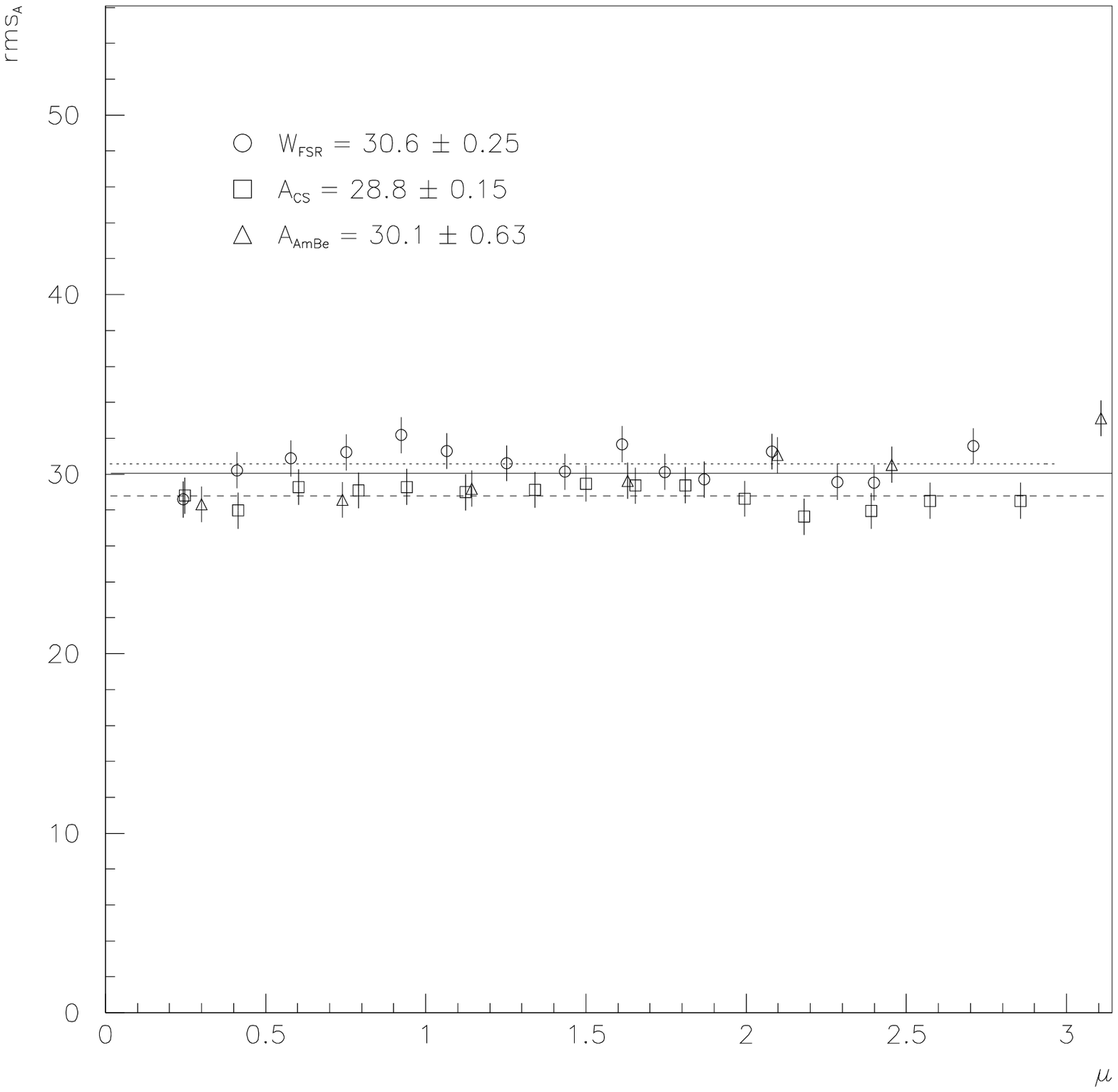}
\par\end{centering}

\caption{\label{fig:PMT_7_RMS}SER width results for PMT $7$ in the ZEPLIN-III
array. The values of $\left(\sigma_{\mathscr{R}}\equiv rms_{A}\right)_{V>V_{thr}}$
and the corresponding averages are shown for: (squares) $^{137}$Cs
data set, (triangles) Am-Be data set and (circles) WIMP-search data
set.}

\end{figure}

\par\end{center}

The array-averaged mean value of the PMT SER $(\left\langle \left\langle \mathscr{R}\right\rangle _{SER}\right\rangle $)
was found to be respectively $\approx5.9\%$ and $\approx10.3\%$
lower for the $^{137}$Cs and Am-Be calibration data sets with respect
to the WIMP-search data set. Simultaneously, the array average of
the relative SER width ($\left\langle \left(\sigma_{\mathscr{R}}/\left\langle \mathscr{R}\right\rangle \right)_{SER}\right\rangle $)
degrade by $\approx2.8\%$ and $\approx8.1\%$ for the $^{137}$Cs
and Am-Be data sets when compared to the WIMP-search data set. It
should be noted that these computations assume a uniform position
distribution of the S1 signals across the entire detector active volume.
This assumption is not exact, especially for the Am-Be data set, due
to the position of the source. Nevertheless, the observed differences
on the mean PMT responses for the different data sets are attributed
to the increase in the resistivity of the bialkali photocathodes at
low temperatures \cite{Araujo:1998aa,Murray:1960aa}. To cope with
this well known effect, the PMTs used have a set of metal tracks deposited
under the photocathode. These tracks decrease the average photocathode
resistivity but also increase its non-uniformity by creating regions
with different abilities to neutralise the charge left by the ejection
of photoelectrons. Thus, depending on the rate, distribution and intensity
of the incoming light pulses, the increase of the resistivity enhances
the local charging of the photocathode which consequently attenuates
and distorts the electric field of the input optics. In addition to
the variation of the quantum efficiency ($\epsilon$) \cite{Araujo:2004mj},
the consequences of this charging are an increase of the variance
of the single photoelectron response and a decrease of the electron
multiplication at the first dynode. The observed qualitative decrease
in the mean response of the PMTs is consistent with the increase in
the rate of energy deposited in the liquid xenon target volume from
the $^{137}$Cs and Am-Be sources and the consequent increase of the
rate of scintillation photons arriving at the photocathodes.

Given the absence of dedicated calibration light sources in the ZEPLIN-III
setup, searching for PMT signals corresponding to thermal single photoelectron
emission ({}``dark counts'') presents the only way to validate the
calibration results obtained using the method described in Sec.~\ref{sec:Methodology}.
For this purpose, a dedicated data set was acquired with the DAQ triggering
from an external pulser ($100$~Hz). The PMT signals were digitized
at $2$~ns sampling over a time segment of $256$~$\mu$s starting
at the trigger instant. The total duration of the run was about $60$~hours
which corresponds to about $500$~s live time for each of the $31$
PMTs. The raw data were reduced using a software threshold of $V_{thr}=3\sigma$
(Eq.~\ref{eq:Threshold}). For each PMT, the spectrum of the pulse
amplitude was used to identify and eliminate the roughly exponential
contribution of the noise just above $V_{thr}$. The surviving pulses
were then assumed to be from thermal single electron emission from
the PMT photocathodes provided that no coincident pulses were found
in any of the other PMT channels. To exclude a connection to any possible
interaction in the xenon target, the anticoincidence was expanded
to all channels during a time window of $200$~ns either side of
the candidate pulse starting time. One can further assume that the
area spectrum of the pulses corresponding to thermal photoelectrons
is a good approximation%
\footnote{See \cite{candy:1985aa,Dossi:2000aa} for further details on reported
differences between the spectra from thermal noise and PMT response
to low intensity light pulses.%
} to the SER of a PMT, given that the probability of having $n>1$
thermal photoelectrons ejected during a time window of $\lesssim100$~ns
is in fact very small. Fig.~\ref{fig:PMT7_SER} shows the pulse area
spectrum of thermal single photoelectron signals from the PMT also
represented in Figs.~\ref{fig:PMT_7} and \ref{fig:PMT_7_RMS}. The
average mean response of the PMTs to single photoelectron signals,
characterized by the mean values of the area spectra, were found to
differ by only $\approx5.3$\% from the values calculated using the
method described in Sec.~\ref{sec:Methodology} using the WIMP-search
data set. The average width of the PMTs SER, characterized by the
\textit{root mean square} of the area spectra, differs $\approx15$\%
from the values estimated using Eq.~\ref{eq:Rel_Var} for the WIMP-search
data set.

\begin{center}
\begin{figure}[p]
\begin{centering}
\includegraphics[width=9.5cm]{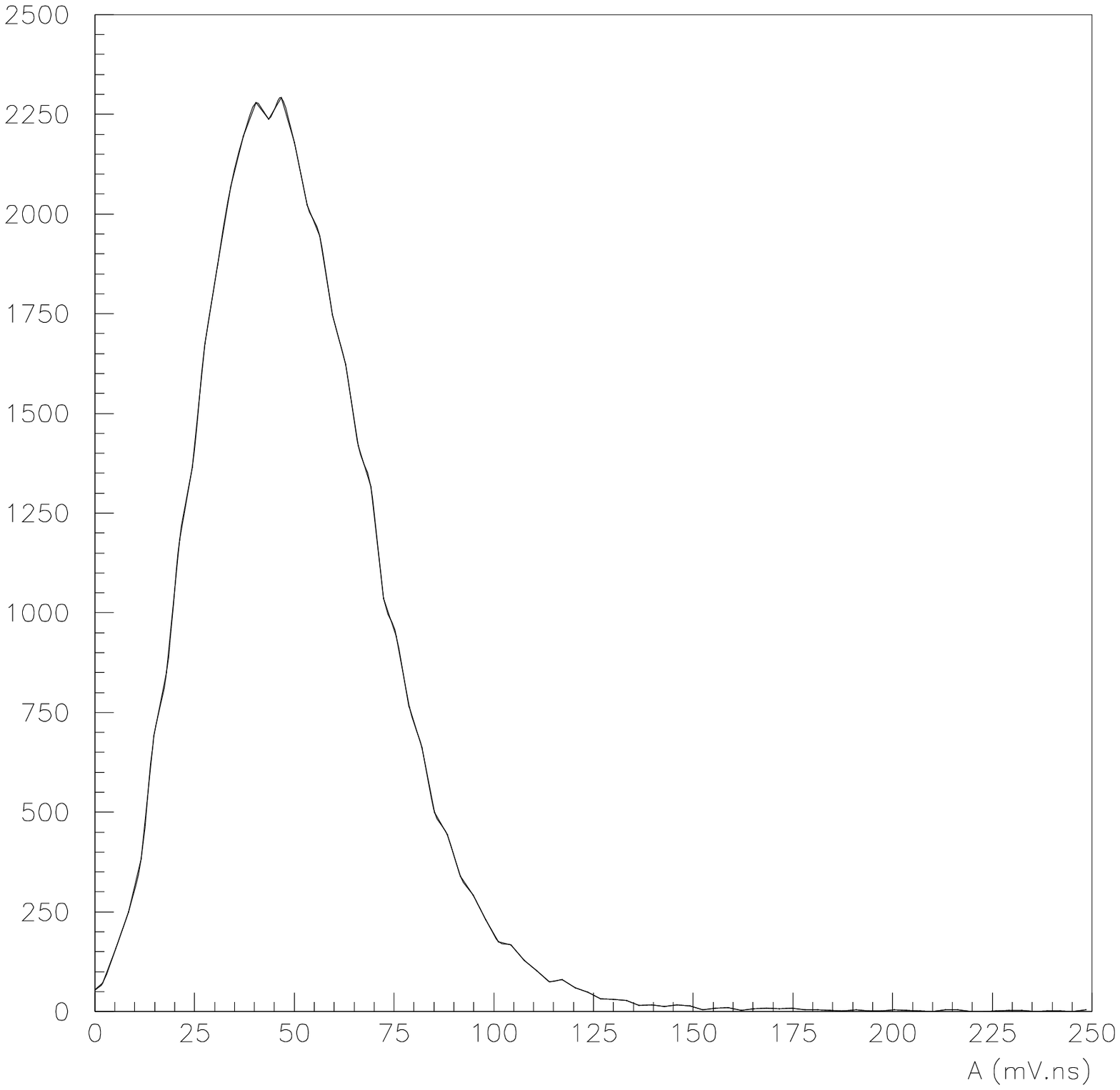}
\par\end{centering}

\caption{\label{fig:PMT7_SER}Pulse area spectrum of thermal single photoelectrons
(for the same PMT also represented in Figs.~\ref{fig:PMT_7} and
\ref{fig:PMT_7_RMS}).}

\end{figure}

\par\end{center}

\section{Conclusions}

In the present work a method to calibrate the SER and assess the linearity
of response of an array of PMTs is described. The method, which does
not require dedicated runs, was applied to the science data from the
ZEPLIN-III experiment. Excellent agreement were found when comparing
the SER mean and width with those derived from a more traditional
measurement using thermal photoelectron emission. Significantly, as
the presented calculations rely only on the statistical description
of the light pulses arriving at the PMTs photocathodes, the method
is suitable to use with any array of photo detectors (e.g. PMTs, APDs,
MPPCs) in applications ranging from low energy rare events searches
to medical PET.

\section*{References}

\bibliographystyle{iopart-num}
\bibliography{bibliography}

\end{document}